\documentclass[letterpaper]{article}
\usepackage[preprint]{aaai2027}
\usepackage[hyphens]{url}
\usepackage{graphicx}
\usepackage{natbib}
\usepackage{caption}
\usepackage{booktabs}
\usepackage{amsmath,amssymb}
\title{Motif-Vocab: Statistically Calibrated Transcription-Factor-Identity Tokenization for Genomic Language Models}
\author{
    Liangyu Li,
    Michael A. White\corresponding
}
\affiliations{
    Washington University in St. Louis\\
    \{l.liangyu, mawhite\}@wustl.edu
}

\begin{document}
\maketitle

\begin{abstract}
Tokenization is a central design choice in genomic language models, yet most
deoxyribonucleic acid (DNA) tokenizers use characters, fixed-length $k$-mers, or frequency-derived
subwords without explicitly using prior information about the specificity of
DNA-binding regulatory factors. We introduce Motif-Vocab, a biologically
informed tokenizer that scans both DNA strands for
statistically calibrated motif matches, emits transcription-factor (TF)
identity tokens, and applies nucleotide, $k$-mer, or byte-pair encoding (BPE)
to unmatched sequence. Motif-specific null distributions put position-weight
matrices (PWMs) of different lengths and degeneracy on a common significance scale;
deterministic overlap rules make the representation reproducible. In
controlled Bidirectional Encoder Representations from Transformers (BERT)
pretraining on two billion base pairs, real motif libraries
outperform randomized-motif controls on 54 of 55 in-scope downstream tasks.
On a
motif-disjoint recognition task derived from DART-Eval Task 2, TF-specific
tokens improve macro-F1 by 0.040 over a position-matched generic motif token
and by 0.033 over a matched no-motif tokenizer (95\% bootstrap confidence
interval: 0.027--0.038). Motif tokens also receive stronger attribution and
produce larger occlusion effects than shuffled controls. Dense no-motif
tokenizers remain strong general-purpose baselines, including a near-tie on
the five-task BERT-base panel. Thus, Motif-Vocab is not a universal accuracy
replacement; it is a targeted, interpretable inductive bias for
motif-sensitive genomic modeling.
\end{abstract}

\section{Introduction}

Tokenization determines the units on which a language model allocates
embeddings, attention, and prediction targets. Natural-language processing
(NLP) commonly uses data-driven subwords such as byte-pair encoding (BPE)
\citep{sennrich2016bpe,kudo2018sentencepiece}; Bidirectional Encoder
Representations from Transformers (BERT) then learns contextual
representations by masked language modeling (MLM) over these units
\citep{devlin2019bert}. Tokenization schemes from NLP are commonly used for deoxyribonucleic
acid (DNA) sequences, where there are only four canonical characters but no spaces or
agreed word boundaries. Existing genomic language models use single bases,
fixed $k$-mers, BPE, or learned segmentation
\citep{ji2021dnabert,zhou2024dnabert2,nguyen2023hyenadna,
schiff2024caduceus,dallatorre2025nucleotide,qiao2024mxdna,
li2026mergedna,huang2026evolen}. These choices balance sequence length and
vocabulary size, but tokens are not chosen to align with meaningful
genomic sequence elements.

This mismatch is especially relevant in regulatory DNA, the non-protein-coding
sequences that control the spatial and temporal expression of genes. Regulatory DNA functions by encoding binding sites for transcription factor (TF) proteins that recognize short, degenerate DNA patterns called motifs. Motifs are typically 6-12 base pairs (bp) in length, can occur on either strand of the DNA double helix, and can overlap. A generic tokenizer may split one motif across tokens, merge it with unrelated context, or represent occurrences of the same TF inconsistently. An attractive alternative to generic tokenizers would be a biologically-motivated tokenization scheme that accounts for TF binding motifs. Motifs are represented as position-frequency matrices (PFMs), which record nucleotide counts at each motif position; adding pseudocounts and normalizing the four nucleotide counts at each position yields position probability matrices (PPMs), which give the probability of each of the four DNA nucleotides occurring at each motif position, or as position-weight matrices (PWMs), which convert those frequencies into position-specific log-odds scores. Open, curated databases such as JASPAR  \citep{baydar2026jaspar} and The Homo sapiens Comprehensive Model Collection (HOCOMOCO) \citep{vorontsov2024hocomoco} collect thousands of PWMs that could be used as prior information for motif-based tokenization. However, motif scanning algorithms cannot be easily incorporated into a tokenization pipeline because a raw motif scanner returns scores that are not comparable across motifs of different lengths and information content, and because de novo scans are too computationally expensive for a general-purpose genomic language model tokenizer.

\begin{figure*}[!t]
\centering
\includegraphics[width=\textwidth]{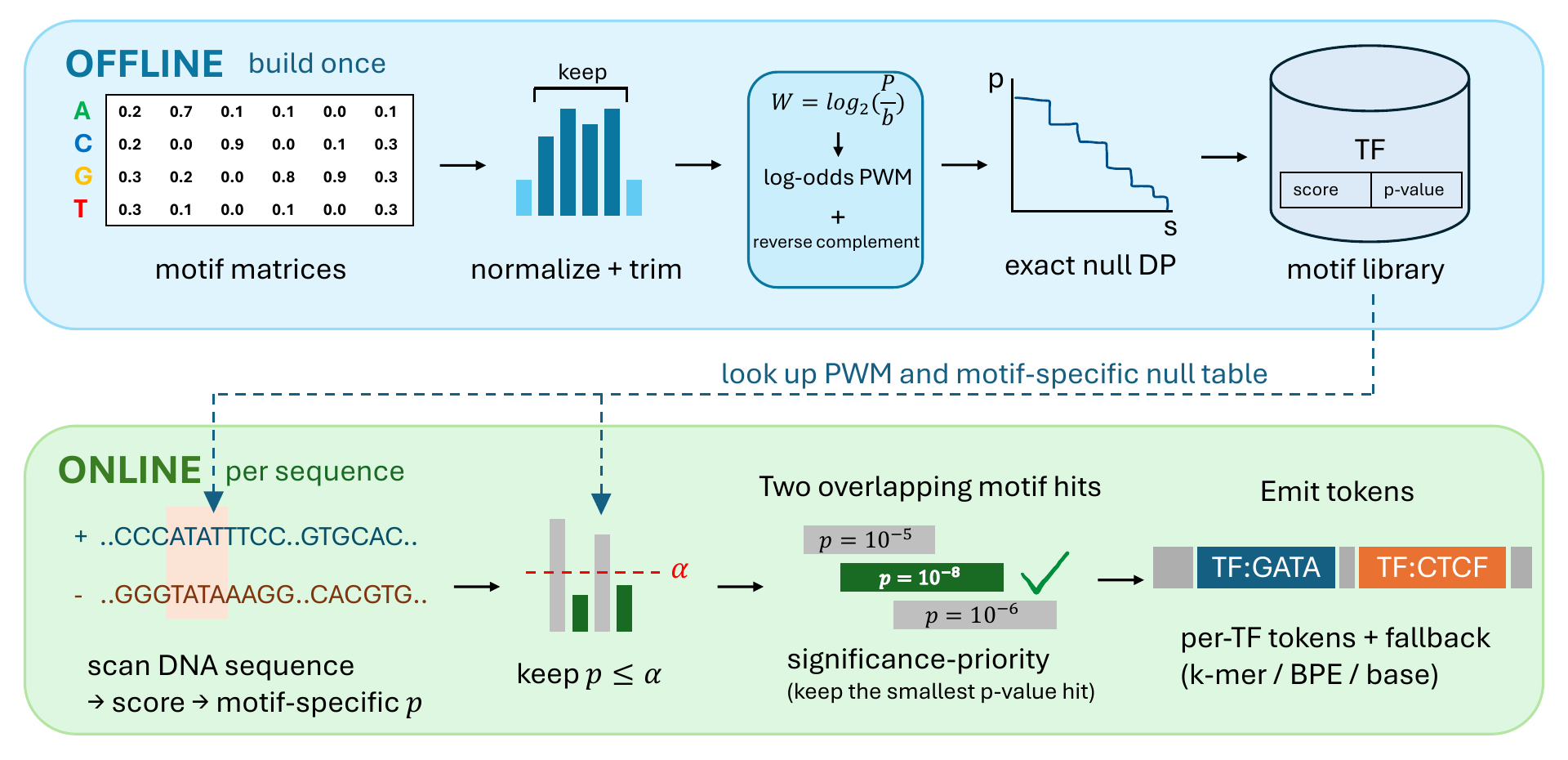}
\caption{\textbf{Motif-Vocab pipeline.} Offline, JASPAR position-frequency
matrices (PFMs) are normalized into PPMs, trimmed only at low-information flanks, converted to
log-odds PWMs, augmented with reverse complements, and assigned exact null
score tables. Online, both strands are scanned; candidates with calibrated
$p\leq\alpha$ become TF-identity tokens, and unmatched sequence uses a
nucleotide, $k$-mer, or BPE fallback. Token spans are retained for base-span
masking and interpretation.}
\label{fig:pipeline}
\end{figure*}

We address these challenges with a DNA tokenizer that identifies statistically significant motifs, paired with a fallback representation to handle sequences between motifs. We propose Motif-Vocab
(Figure~\ref{fig:pipeline}).
It calibrates each motif against a background-specific null distribution,
maps significant matches from all matrices and both strands to one token per
TF, resolves competing matches deterministically, and fills unmatched regions
with bases, $k$-mers, or BPE. This creates an input interface in which a token
can be traced to a known regulatory factor. We can then test whether the model
uses that token through attribution, which assigns prediction importance to
input tokens, and occlusion, which masks a token and measures the prediction
change.

We find that Motif-Vocab does not universally supersede generic
tokenizers, but it provides an interpretable inductive bias for motif-sensitive
regulatory tasks. We match the internal encoder, pretraining base-pair corpus,
downstream splits, and nominal training budget, while varying tokenization. We
compare real motifs with
no-motif, randomized-motif, and position-matched generic-token controls, then
add public DNABERT-2, Nucleotide Transformer version 2 (NT-v2), and Caduceus
baselines. The
evidence supports three contributions:
\begin{itemize}
  \item We introduce a deterministic TF-identity tokenizer with motif-specific
  significance calibration, both-strand scanning, nucleotide, 6-mer, or BPE fallback units,
  and overlap-aware MLM.
  \item We provide a controlled study spanning 12 tokenizers variants, 55 in-scope
  downstream tasks, three
  seeds, BERT-base scale-up, public DNA language models, and a motif-disjoint
  DART-Eval-derived evaluation.
  \item We show that real TF identity contributes beyond segmentation geometry
  and supports mechanistic interpretation on motif-sensitive tasks, while
  clearly delimiting settings where dense no-motif tokenizers remain stronger on motif-independent tasks.
\end{itemize}

\section{Background and Related Work}

\paragraph{Tokenization and domain priors.}
BPE and SentencePiece compress frequent substrings, but segmentation quality
depends on the domain and objective
\citep{bostrom2020bpe,rust2021tokenizer}. Byte-level models remove learned
boundaries at the cost of longer sequences \citep{xue2022byt5}. Motif-Vocab
instead retains generic fallback units while reserving atomic tokens for
detected motifs.

\paragraph{Language models and AI for Science.}
Large language models (LLMs) and foundation models show that pretraining on
large corpora can produce transferable representations
\citep{brown2020language,bommasani2021foundation}. Multimodal LLMs (MLLMs)
extend this idea by aligning language with another modality through contrastive
pretraining or cross-modal conditioning
\citep{radford2021clip,alayrac2022flamingo}. A recurring theme in AI for
Science is that raw measurements are not enough: useful models often encode
domain structure in the representation, objective, or simulator
\citep{wang2023scientificdiscovery}. Examples include physics-informed losses,
message-passing simulators, neural operators, and scientific MLLM systems
\citep{raissi2019pinn,sanchezgonzalez2020gns,kovachki2023neuraloperator,huang2025peace}.
Our model is not itself a multimodal LLM. The relevance is representational:
MLLMs connect raw observations to a second semantic modality, whereas
Motif-Vocab connects DNA substrings to symbols from a curated TF knowledge
source. The TF token stream can therefore act as a lightweight bridge between
sequence and biological semantics, and could later support DNA--text or
DNA--knowledge MLLMs. Here we isolate the preceding question: whether that
symbolic alignment is already useful within a sequence-only encoder.

\paragraph{Genomic language models and tokenizers.}
DNABERT uses overlapping $k$-mers, DNABERT-2 uses BPE, and Nucleotide
Transformer uses fixed $k$-mers at larger multi-species scale
\citep{ji2021dnabert,zhou2024dnabert2,dallatorre2025nucleotide}. HyenaDNA and
Caduceus operate at nucleotide resolution with long-context sequence
architectures \citep{nguyen2023hyenadna,schiff2024caduceus}. Recent work makes
segmentation itself adaptive: MxDNA learns overlapping and discontinuous units,
MergeDNA merges context-dependent tokens, and DNAChunker learns variable-length
chunks \citep{qiao2024mxdna,li2026mergedna,kim2026dnachunker}. EvoLen biases BPE
vocabulary construction with evolutionary conservation
\citep{huang2026evolen}. Closest to our aim, DNAMotifTokenizer injects motif
knowledge and motivates our low-information flank trimming
\citep{zhou2025dnamotiftokenizer}. DNAMotifTokenizer is closest in using motif knowledge. Motif-Vocab differs in four central respects: it preserves full PWM degeneracy, calibrates each motif on a common p-value scale, collapses multiple matrices and strands to one TF token, and uses deterministic overlap resolution with base-span masking \citep{zhou2025dnamotiftokenizer}.

\paragraph{Regulatory sequence models.}
Predicting regulatory function from DNA sequence alone has long been an important modeling goal in genomics\citep{beer2004predicting}.  Task-specific neural networks, such as DeepSEA\citep{zhou2015deepsea}, Basenji\citep{kelley2018basenji}, BPNet\citep{avsec2021bpnet}, and Enformer\citep{avsec2021enformer}, have shown some success at predicting regulatory DNA function, the impact of DNA sequence variants, and long-range interactions between distant regulatory elements in the genome. These models learn motifs as latent features, which are then extracted through post-hoc attribution methods\citep{avsec2021bpnet,shrikumar2018technical}. With genomic language models, motif recovery is challenging because motifs may span token boundaries, and their latent representations must be mapped back to nucleotide-level patterns. Motif-Vocab instead presents calibrated motif identities at the model input.

\paragraph{Motif scoring and regulatory evaluation.}
A PWM assigns a log-odds score to each base at each motif position
\citep{stormo2013specificity}. Find Individual Motif Occurrences (FIMO)
established the value of calibrated motif occurrence tests
\citep{grant2011fimo}; we adapt this idea to token generation
and precompute exact score distributions. Our matrices come from JASPAR CORE
2026 \citep{baydar2026jaspar}. For evaluation, Genomic Benchmarks provides
reproducible regulatory sequence classification tasks
\citep{gresova2023genomicbenchmarks}; DNABERT-2 and Nucleotide Transformer add
broader genome-understanding panels; and DART-Eval targets regulatory DNA under
zero-shot, probing, and fine-tuning protocols \citep{patel2024darteval}.

\section{Method}

\subsection{Calibrated Motif Matching}

Let $x\in\{A,C,G,T\}^n$ and let $P\in[0,1]^{L\times4}$ be the position
probability matrix of a length-$L$ motif. With background base probabilities
$b_a$, its PWM is $W_{k,a}=\log_2(P_{k,a}/b_a)$ and the score at start $i$ is
\begin{equation}
S_W(x,i)=\sum_{k=1}^{L}W_{k,x_{i+k-1}}.
\label{eq:pwm-score}
\end{equation}
Raw scores cannot be compared across motifs. Under the assumption of independent background
bases $B_k\sim b$, we therefore use
$p_W(s)=\Pr[\sum_k W_{k,B_k}\geq s]$. A dynamic program convolves the
per-position score distributions on a grid of precision $\gamma=100$ and
stores the survival table once per motif. At tokenization time, a score needs
one lookup; a candidate is retained when $p_W(S_W)\leq\alpha$, with
$\alpha=10^{-4}$ in all main experiments. The observed 2-billion-base-pair
training background supplies $b$. This calibration converts
motif-specific raw scores into a common significance scale: a short,
low-degeneracy motif and a longer, more degenerate motif are compared by how
unlikely their scores are under the same background, not by the raw score
magnitude.

We add pseudocount $0.01$, normalize each PFM to obtain PPM, and
trim only consecutive maximum PPM probability of flank columns below the information threshold $0.4$; the
motif interior is never removed. The trimming rule is inspired by the
motif-boundary refinement in DNAMotifTokenizer
\citep{zhou2025dnamotiftokenizer}. We then build forward and reverse-complement
PWMs. The offline library stores the TF name, matrix accession, strand, trimmed
length, PWM, and calibrated survival table. JASPAR matrix accessions sharing
the same exact TF name map to one token, so versions or alternative matrices
for GATA1 map to $\langle\mathrm{TF:GATA1}\rangle$. Related TFs, such as GATA1 and GATA2, remain
distinct, even though they may have highly similar motifs. Reverse-complement hits
also retain the same TF identity while strand is stored as metadata.

\subsection{TF Vocabulary, Overlap, and Masking}

\begin{figure*}[t]
\begin{minipage}[t]{0.48\textwidth}
\centering
\includegraphics[width=\linewidth]{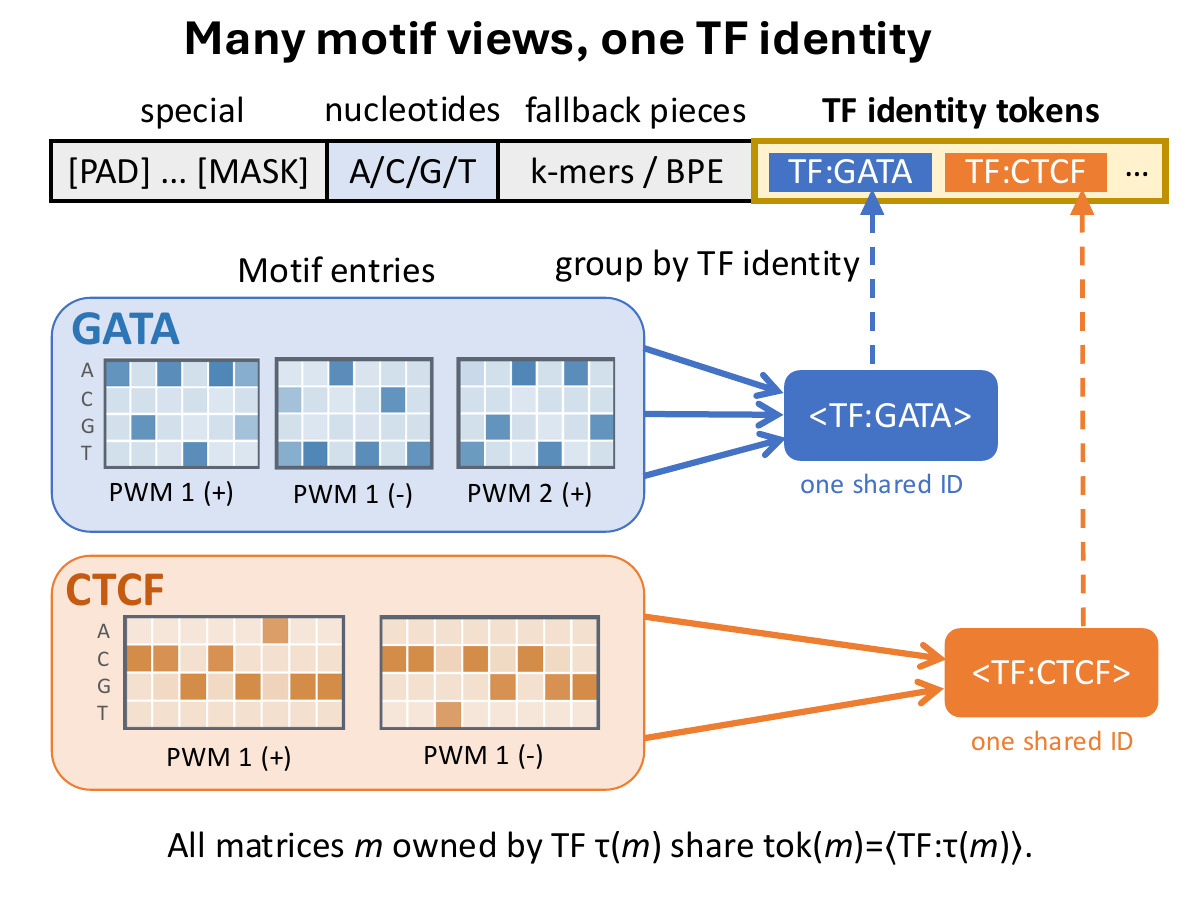}
\caption{\textbf{TF-level vocabulary.} Multiple matrices and both strands
map to one exact TF identity token. Different GATA factors remain separate;
the schematic label GATA denotes one example identity, not a merged TF family.
Fixed special and nucleotide IDs and fallback units remain available for
unmatched sequence.}
\label{fig:tf-vocab}
\end{minipage}\hfill
\begin{minipage}[t]{0.48\textwidth}
\centering
\includegraphics[width=\linewidth]{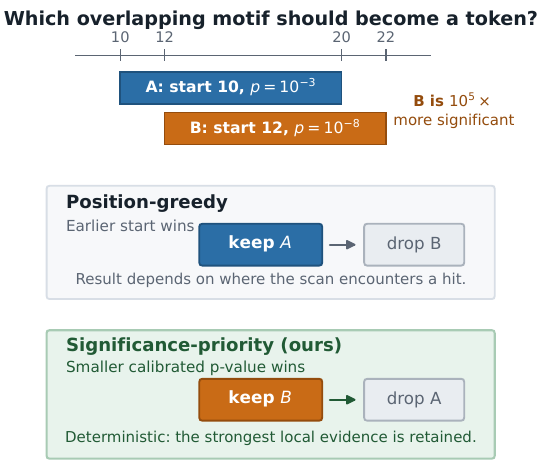}
\caption{\textbf{Non-overlap conflict resolution.} An earlier weak hit can
hide a later strong hit under position-greedy segmentation. Significance
priority retains the smaller $p$-value. Coupled-overlap variants instead emit
the best hit at each start and do not apply this exclusion step.}
\label{fig:overlap}
\end{minipage}
\end{figure*}

The vocabulary contains five special tokens, the four bases, the fallback
inventory, and one token $\langle\mathrm{TF}:t\rangle$ per TF
(Figure~\ref{fig:tf-vocab}). The final library uses all non-redundant
Vertebrata entries in JASPAR CORE 2026, rather than a human/mouse-only filter:
1,019 forward JASPAR PWMs, their reverse complements, collapsing accessions by TF name yields 964 unique TF identities and therefore 964 TF token IDs; strand orientation does not create an additional token. The
matrix evidence is annotated to \emph{Homo sapiens}, \emph{Mus musculus},
\emph{Rattus norvegicus}, \emph{Rattus rattus}, \emph{Oryctolagus cuniculus},
\emph{Gallus gallus}, and \emph{Xenopus laevis}; one PFM lacks a species field,
and one has four source species. These species fields record the origin of the
evidence used to construct a profile, whereas Vertebrata is the profile's
taxonomic scope \citep{baydar2026jaspar}. Each emitted token carries
its half-open base span $[s,e)$, strand, motif accession, score, and $p$-value.
For an input sequence, Motif-Vocab scans every start position on both strands.
Each significant hit is represented as
$h=(s_h,e_h,t_h,r_h,S_h,p_h)$, where $s_h,e_h$ are base coordinates,
$t_h$ is the TF identity, $r_h$ is the strand, and $S_h,p_h$ are the PWM score
and calibrated significance. Uncovered bases are encoded by the selected
fallback tokenizer: nucleotide, fixed $k$-mer, or BPE. This makes the method a
drop-in tokenizer rather than a separate feature extractor.

Motif-Vocab supports two regimes. In \emph{non-overlap} mode, all significant
hits are sorted by
\begin{equation}
\kappa(h)=(p_h,-S_h,-\ell_h,s_h,\mathrm{name}_h),
\label{eq:priority}
\end{equation}
and are greedily accepted if they do not overlap an accepted interval. This keeps
the most significant local evidence and gives deterministic ties
(Figure~\ref{fig:overlap}). In \emph{coupled-overlap} mode, used by K6-OV
(overlapping 6-mer fallback) and BPE-OV (overlapping byte-pair-encoding
fallback), the best candidate at each start is emitted and the scan
advances one base; fallback $k$-mers or BPE pieces also advance one base.
Thus motif and fallback evidence can cover the same bases. Because ordinary
token MLM would over-mask or leak overlapping content, we
sample masked base spans and mask every token intersecting those spans; the
prediction target is therefore tied to genomic coordinates rather than to an
arbitrary segmentation.

\paragraph{Variant notation.}
The variant labels describe tokenizer conditions, not different internal
encoder architectures: all broad controlled variants use the same BERT-small
model. A label combines the motif layer, fallback tokenizer, and overlap mode;
for example, Motif K6-OV means real JASPAR TF tokens plus an overlapping 6-mer
fallback. For the shuffled-motif control, we randomly permuted the motif positions for each JASPAR PFM and then independently shuffled the A/C/G/T count labels within each position. This preserves motif length and the per-position count distributions while destroying true TF sequence specificity, thereby controlling for generic motif-like structure rather than TF identity. Generic identity (Generic ID) is a separate DART-T2 control that retains real hit spans
but maps every TF to one shared \texttt{<MOTIF>} token.
\begin{center}
{\small
\setlength{\tabcolsep}{2.6pt}
\begin{tabular}{@{}p{0.20\columnwidth}p{0.31\columnwidth}p{0.41\columnwidth}@{}}
\toprule
Label part & Values & Meaning \\
\midrule
Motif layer & No motif; Motif; Shuffled & none; real JASPAR; randomized matrices \\
Fallback & nucleotide; K6; BPE & single base; fixed 6-mer; byte-pair encoding \\
Suffix & NO; OV & non-overlap; coupled one-base overlap \\
Scale & BERT-small; BERT-base & controlled main study; selected scale-up \\
\bottomrule
\end{tabular}}
\end{center}

For $M$ motifs with total length $L_{\rm tot}=\sum_m L_m$ and $H$ retained
candidates, online scanning and non-overlap resolution cost
$O(nL_{\rm tot}+H\log H)$; coupled overlap omits the global sort. Offline null
construction costs $O(\gamma\Delta\sum_m L_m^2)$ for score range $\Delta$ and
is paid once. In practice, the expensive step is library construction, while
training and fine-tuning reuse serialized motif tables and tokenized caches.
Serialized libraries and tokenizer metadata make the mapping deterministic and
reusable.

\section{Experiments}

\subsection{Data, Models, and Benchmarks}

\paragraph{Pretraining.}
We use non-overlapping 512-base-pair (bp) windows from the GENCODE Release 49
human GRCh38.p14 and Release M38 mouse GRCm39 primary assemblies
\citep{mudge2025gencode}. We retain canonical autosomes and chromosome X,
reserve chromosome 8 from both species for validation and chromosome 9 for
test, and use the remaining canonical chromosomes for training. Windows with
more than 5\% characters other than A, C, G, or T are discarded. We
prespecified 2.00 billion
unique training bp as a controlled-compute budget: it keeps DNA exposure equal
across 12 tokenizer variants and three seeds while making the full comparison
tractable in compute, tokenized-cache storage, and wall time. The cap is an
experimental budget, not a biological boundary or a claim that 2 billion bp
represent both complete genomes; processing every eligible base from both
assemblies for every variant would increase cost without improving isolation
of the tokenizer variable. We allocate 3,906,250 training windows
proportionally to the eligible window count of each species and chromosome,
then retain windows in genomic order up to each quota. This is deterministic
quota-based truncation, not random sampling. The resulting 2.00
billion bp comprise 1.067 billion human and 0.933 billion mouse bp; validation
and test contain 0.270 billion and 0.243 billion bp. Thus the corpus is a reproducible subset of the
two assemblies, not their full combined sequence. BPE is trained only on the
training split. Each three-epoch BERT-small run therefore receives 6.00 billion
bp of cumulative sequence exposure from 2.00 billion unique training bp.
Controlled BERT-small models have
6 layers, width 512, 8 attention heads, and feed-forward width 2,048; they use
15\% MLM, three corpus epochs, and the same input base-pair budget. BERT-base scale-up
uses 12 layers, width 768, 12 heads, and width 3,072. Tokenization necessarily
changes sequence length, vocabulary-dependent embedding parameters, and
masking implementation; these are measured consequences of the tokenizer,
not held-fixed quantities. All downstream summaries average seeds 13, 42, and
3407 and report macro-F1.

\paragraph{Broad and core panels.}
The broad controlled study reports 55 binary or multiclass tasks: six
Genomic Benchmarks (GB), 34 Genome Understanding Evaluation (GUE) tasks from
DNABERT-2, and 15 Nucleotide Transformer (NT) tasks. They cover promoters,
enhancers, open chromatin regions (OCRs), splice sites, histone marks, and TF
binding. Many broad tasks are only indirectly connected to JASPAR TF-binding
motifs, so they should be read as generalization stress tests rather than
tasks designed to favor Motif-Vocab. We uniformly exclude three initially
acquired GUE task groups: phage fragment classification, virus/coronavirus
disease 2019 (COVID-19) classification, and virus species classification. Their taxonomic sequence
labels do not directly test eukaryotic TF binding or the cis-regulatory
mechanisms represented by the JASPAR vocabulary. JASPAR CORE is a eukaryotic TF
motif collection and does not supply a viral TF-binding vocabulary for these
labels. This semantic scope rule is
applied to every tokenizer, independent of score. Twelve
tokenizer variants and three seeds therefore give 1,980 paper-facing runs. All
2,088 original runs remain in provenance. An exploratory five-task core panel tests scale
transfer on human non-TATA promoters, human OCRs, synthetic motif disruption,
enhancers, and TATA-box promoters. The core panel was chosen for
endpoint diversity and closer TF-regulatory relevance, not as a prespecified
universal benchmark.

\paragraph{DART-Eval.}
DART-Eval defines five regulatory evaluations: (T1) regulatory sequence versus
dinucleotide-matched control; (T2) sensitivity to 1,443 TF motifs; (T3)
cell-type-specific regulatory DNA; (T4) quantitative regulatory activity; and
(T5) genetic-variant effect prediction \citep{patel2024darteval}. We use the
released T2 positive and shuffled motif-insertion sequences, generated from
the HOCOMOCO v12 motif library
\citep{vorontsov2024hocomoco}, to construct a
balanced supervised recognition stress test (100k/20k/20k train/validation/
test), with all 1,443 motif identities disjoint across splits. We call it
\emph{DART-T2 motif recognition}; ``motif-disjoint'' means that the motif
identities used to generate positives in the held-out split are absent from
fine-tuning. It tests whether a tokenizer transfers to unseen motif identities
under one fine-tuning protocol; it is not the official DART zero-shot
likelihood/embedding score. Project-specific T5 diagnostics are kept in the
supplement because their protocol and objective differ from the official
benchmark.

\begin{figure*}[t]
\centering
\includegraphics[width=\textwidth]{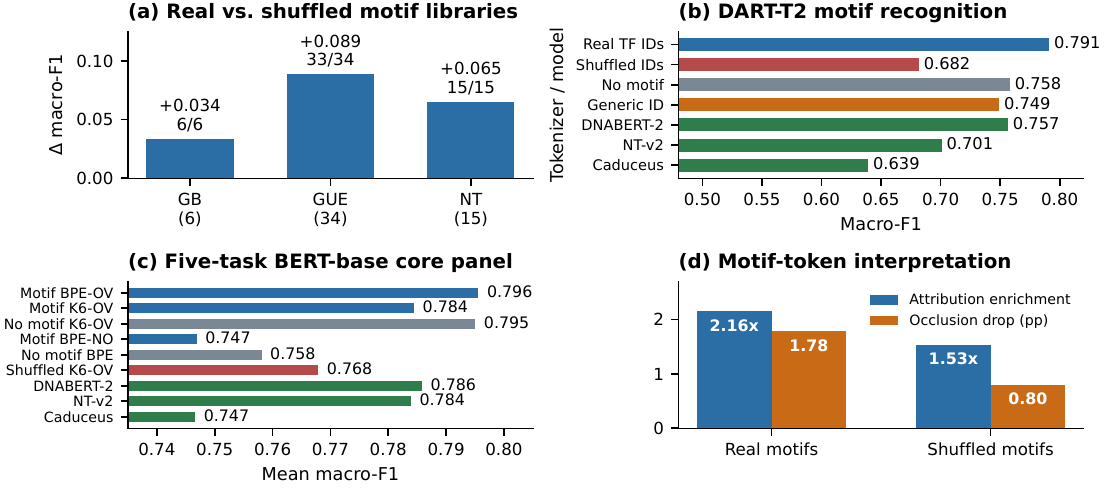}
\caption{\textbf{Main empirical evidence.} (a) Real motifs consistently beat
randomized libraries across broad families, although no-motif K6 is the best
broad aggregate. (b) On DART-T2 recognition, real TF tokens beat a
shuffled-motif control, position-matched generic token, matched no-motif
control, and evaluated public models. The four internal conditions in panel
(b) all use K6-OV, and the vertical axis lists tokenizer or model conditions.
(c) Five-task BERT-base means show a practical tie between Motif BPE-OV and
No-motif K6-OV and include non-overlap BPE controls. (d) Real motif tokens
receive stronger attribution and cause a larger probability drop when
occluded than shuffled tokens.}
\label{fig:main-results}
\end{figure*}

\begin{table*}[!t]
\centering
\footnotesize
\setlength{\tabcolsep}{2.6pt}
\begin{tabular}{@{}llrrrrrrr@{}}
\toprule
Group & Task / aggregate & \shortstack{Motif\\BPE-OV} & \shortstack{Motif\\K6-OV} & \shortstack{No motif\\K6-OV} & \shortstack{Shuffled\\K6-OV} & DNABERT-2 & NT-v2 & Caduceus \\
\midrule
Broad & GB (6) & 0.839 & \underline{0.845} & \textbf{0.856} & 0.811 & -- & -- & -- \\
 & GUE (34) & 0.719 & \underline{0.732} & \textbf{0.744} & 0.644 & -- & -- & -- \\
 & NT (15) & 0.767 & \underline{0.773} & \textbf{0.781} & 0.708 & -- & -- & -- \\
 & Family-balanced & 0.775 & \underline{0.784} & \textbf{0.794} & 0.721 & -- & -- & -- \\
 & All 55 tasks & 0.745 & \underline{0.756} & \textbf{0.766} & 0.680 & -- & -- & -- \\
\midrule
Core & Non-TATA promoter & \underline{0.920} & 0.919 & 0.915 & 0.906 & \textbf{0.930} & 0.898 & 0.864 \\
 & Open chromatin & \underline{0.798} & 0.793 & 0.786 & 0.776 & \textbf{0.807} & 0.785 & 0.750 \\
 & Motif disruption & 0.538 & \textbf{0.551} & 0.522 & 0.518 & 0.496 & \underline{0.544} & 0.524 \\
 & Enhancer & \textbf{0.803} & 0.792 & \underline{0.797} & 0.768 & 0.758 & 0.759 & 0.736 \\
 & TATA promoter & 0.920 & 0.868 & \textbf{0.955} & 0.871 & \underline{0.939} & 0.934 & 0.860 \\
\midrule
DART-derived & DART-T2 motif recognition & \underline{0.773} & \textbf{0.791} & 0.758 & 0.682 & 0.757 & 0.701 & 0.639 \\
\bottomrule
\end{tabular}
\begin{minipage}{0.99\textwidth}
\scriptsize \textit{Notes.} BPE is byte-pair encoding; K6-OV is an overlapping 6-mer fallback. DART-T2 recognition is our supervised motif-disjoint derivative, not the official zero-shot DART score. The shuffled value uses validation-selected checkpoints; all three held-out runs pass prediction-health gates. External models provide context rather than data-matched tokenizer ablations.
\end{minipage}
\caption{Comprehensive three-seed macro-F1 comparison. Broad results use common 55-task non-virus support. Three GUE phage/virus classification tasks are outside the eukaryotic regulatory scope represented by the JASPAR transcription-factor vocabulary and are excluded uniformly. Family-balanced weights GB, GUE, and NT equally. \textbf{Bold} indicates the best result and \underline{underlining} indicates the second-best valid result in each row.}
\label{tab:main-comprehensive}
\end{table*}

\begin{table}[!t]
\centering
\footnotesize
\setlength{\tabcolsep}{2.2pt}
\begin{tabular}{@{}lrrrr@{}}
\toprule
Tokenizer & GB & GUE & NT & All \\
\midrule
No motif K6-OV & \textbf{0.856} & \textbf{0.744} & \textbf{0.781} & \textbf{0.766} \\
No motif nucleotide & 0.849 & 0.735 & \underline{0.774} & \underline{0.758} \\
No motif BPE & \underline{0.854} & \underline{0.742} & 0.750 & 0.756 \\
Motif K6-OV & 0.845 & 0.732 & 0.773 & 0.756 \\
No motif K6-NO & 0.852 & 0.722 & 0.773 & 0.750 \\
Motif BPE-OV & 0.839 & 0.719 & 0.767 & 0.745 \\
Motif BPE-NO & 0.833 & 0.724 & 0.727 & 0.737 \\
Motif K6-NO & 0.824 & 0.711 & 0.720 & 0.726 \\
Shuffled BPE-OV & 0.834 & 0.686 & 0.755 & 0.721 \\
Shuffled BPE-NO & 0.825 & 0.697 & 0.721 & 0.717 \\
Shuffled K6-NO & 0.823 & 0.686 & 0.715 & 0.709 \\
Shuffled K6-OV & 0.811 & 0.644 & 0.708 & 0.680 \\
\bottomrule
\end{tabular}
\caption{Controlled tokenizer comparison on common 55-task non-virus support. GB, GUE, and NT contain 6, 34, and 15 tasks; All is the task-level macro average. OV and NO denote overlapping and non-overlapping tokenization. Bold and underlining mark the best and second-best result in each column.}
\label{tab:main-broad-all}
\end{table}

\begin{table}[!t]
\centering
\footnotesize
\setlength{\tabcolsep}{1.2pt}
\begin{tabular}{@{}lrrp{0.35\columnwidth}@{}}
\toprule
Evidence & Real motif & Control & Effect / support \\
\midrule
Broad GB & 0.845 & 0.811 & +0.034; 6/6 tasks \\
Broad GUE & 0.732 & 0.644 & +0.089; 33/34 tasks \\
Broad NT & 0.773 & 0.708 & +0.065; 15/15 tasks \\
\midrule
DART generic ID & 0.790 & 0.749 & +0.040; 3/3 seeds \\
DART no motif & 0.791 & 0.758 & \mbox{+0.033\,[0.027,0.038]} \\
DART shuffled & 0.791 & 0.682 & \mbox{+0.109\,[0.095,0.129]} \\
\midrule
Calibration & 0.858 & 0.612 & +0.245 \\
Attribution enrich. & 2.157 & 1.533 & 1.41$\times$ \\
Motif occlusion & 0.0178 & 0.0080 & 2.22$\times$ \\
\bottomrule
\end{tabular}
\begin{minipage}{0.98\columnwidth}
\scriptsize Real motif is the real JASPAR TF-identity condition for accuracy rows and the motif-specific signal for diagnostic rows. Control is the corresponding randomized, generic, no-motif, or raw-score comparison. The final three rows are diagnostics, not macro-F1.
\end{minipage}
\caption{Controlled evidence for biological motif structure and TF-level token utility. Broad rows compare real and randomized K6-OV libraries; DART-T2 controls preserve fallback segmentation and, for generic identity, motif hit positions. Accuracy rows report macro-F1; the final three diagnostic rows report calibration, attribution enrichment, and occlusion effects.}
\label{tab:main-ablation}
\end{table}

\subsection{Results and Ablations}

\paragraph{Broad performance scopes the claim.}
Tables~\ref{tab:main-comprehensive} and~\ref{tab:main-broad-all} show that
dense overlapping 6-mers are the
strongest general-purpose tokenizer: on common 55-task non-virus support,
No-motif K6-OV obtains 0.766 overall and 0.794 when the three families are
weighted equally, versus 0.756/0.784 for Motif K6-OV. The lower broad-task aggregate of Motif K6-OV shows that Motif-Vocab is not a universal accuracy replacement. Most broad tasks do not directly ask whether a
JASPAR TF motif is present. Their labels can depend on general sequence
composition, splice-site syntax, histone or epigenetic state, or regulatory
context that is only partly explained by TF-binding specificity. Dense K6-OV
supplies a 6-mer at every sequence start, preserving uniform local sequence
evidence for such endpoints. Motif-Vocab is deliberately narrower: at a
significant match it substitutes a categorical TF identity for one fallback
unit. Its advantage should therefore be judged against matched
motif controls and on TF-sensitive tasks, not by requiring it to dominate a
general benchmark average. The matched real-versus-randomized comparison is different: real
Motif K6-OV improves GB/GUE/NT by 0.034/0.089/0.065 and is positive on 54 of
55 task means (two-sided sign test $p=3.11\times10^{-15}$).
The randomized library actually covers more bases (0.788 versus 0.723) and
activates more distinct motifs (1,796 versus 1,658), so this gain is not
explained by greater token density. Because hit positions still differ, this
control tests biological motif structure, not identity alone.

\paragraph{TF identity survives stricter controls.}
The generic-ID model, also called the generic-motif control, uses the same
real-JASPAR hits, spans, K6-OV fallback tokens, pretraining seed, and budget as
Motif K6-OV, but replaces every named TF token with one shared
\texttt{<MOTIF>} token. It therefore preserves where motif-shaped segments
occur and their tokenization geometry while removing which TF matched.
Motif-Vocab reaches
$0.790\pm0.003$ on DART-T2, versus $0.749\pm0.005$ for generic ID, a 0.040
gain in all three seeds. Against the matched no-motif tokenizer, a paired
hierarchical bootstrap over examples and seeds gives 0.033
[0.027, 0.038]. Against the validation-selected shuffled-motif control, the
corresponding gain is 0.109 [0.095, 0.129], with all three held-out runs
passing prediction-health checks. This is the cleanest evidence that TF-level
categorical identity contributes beyond
where a motif-shaped token occurs. Motif-Vocab also exceeds the evaluated
pretrained DNABERT-2, NT-v2, and Caduceus baselines under our supervised
protocol. These external LLMs were pretrained on substantially larger and more
diverse genomic corpora than our internal models trained on 2 billion base
pairs, so the
comparison is intentionally conservative for Motif-Vocab. However, differences in pretraining data and architecture mean that it is not a data-matched causal comparison of tokenizers.

\paragraph{Scale and interpretation.}
On the five-task BERT-base panel, Motif BPE-OV scores 0.7956 and No-motif
K6-OV scores 0.7950: an effective tie, with different task wins
(Figure~\ref{fig:main-results}c). We therefore do not claim a scale-wide
accuracy advantage. The benefit is an interpretable interface: integrated-gradient
attribution \citep{sundararajan2017axiomatic} assigns prediction importance to
tokens, and attribution to real motif tokens is enriched 2.16-fold over the
token baseline, versus 1.53-fold for shuffled controls. Motif-token occlusion
masks motif tokens and measures the predicted-class probability drop; this
drop is 0.0178 for real motifs versus 0.0080 for shuffled controls. Both attribution enrichment and motif occlusion are averaged over seed-42 runs on the same five core tasks. These
independent
diagnostics indicate that the classifier uses real motif identities as
predictive evidence rather than merely benefiting from compression.
Table~\ref{tab:main-ablation} consolidates the controlled accuracy,
calibration, attribution, and occlusion evidence.

\section{Limitations}

TF tokens do not capture binding affinity: after thresholding, different significant
scores for the same TF map to the same identity token, although score and
$p$-value remain in metadata. The prior can help when TF identity is relevant, but it can discard
within-motif nucleotide variation and cannot encode TF abundance, cell state,
or binding strength. Additionally, related TFs often bind nearly identical motifs, and thus in practice token IDs do not uniquely identify the TFs that bind at that genomic site. Independent
background bases make calibration reproducible but do not model local genomic
dependence. Internal experiments match input base-pair data and architecture, yet token
length, vocabulary size, masking, and compute change with tokenization. Public
models are contextual baselines, not data-matched ablations. Finally, DART-T2
recognition is a supervised derivative designed to isolate motif transfer, not
an official DART leaderboard score; the official five-task suite remains a
distinct evaluation target.

\section{Conclusion}

Motif-Vocab turns calibrated prior information about the specificity of DNA-binding regulatory factors into TF-identity tokens while
preserving generic sequence fallback. Controlled comparisons show a bounded
but useful result: dense overlapping $k$-mers remain excellent general
tokenizers, whereas real TF identities provide stable gains over
position-matched generic tokens and support direct attribution on
motif-sensitive recognition. More broadly, tokenization is an underused point
for injecting structured scientific knowledge into foundation models. Future
work could condition motif tokens on cell state and extend them from identification of isolated
sites to learning the interactions between them that make up the grammar of regulatory DNA.

\section*{Generative AI Use Disclosure}

Generative AI tools assisted with language editing, code review and debugging,
and manuscript organization. The authors verified all scientific claims,
references, analyses, and reported values. All experimental measurements were
produced by the documented computational pipeline.

\clearpage
\bibliography{references}

\begin{thebibliography}{39}
\providecommand{\natexlab}[1]{#1}

\bibitem[{Alayrac et~al.(2022)Alayrac, Donahue, Luc, Miech, Barr, Hasson, Lenc,
  Mensch, Millican, Reynolds et~al.}]{alayrac2022flamingo}
Alayrac, J.-B.; Donahue, J.; Luc, P.; Miech, A.; Barr, I.; Hasson, Y.; Lenc,
  K.; Mensch, A.; Millican, K.; Reynolds, M.; et~al. 2022.
\newblock Flamingo: A Visual Language Model for Few-Shot Learning.
\newblock In \emph{Advances in Neural Information Processing Systems},
  volume~35, 23716--23736.

\bibitem[{Avsec et~al.(2021{\natexlab{a}})Avsec, Agarwal, Visentin, Ledsam,
  Grabska-Barwinska, Taylor, Assael, Jumper, Kohli, and
  Kelley}]{avsec2021enformer}
Avsec, {\v{Z}}.; Agarwal, V.; Visentin, D.; Ledsam, J.~R.; Grabska-Barwinska,
  A.; Taylor, K.~R.; Assael, Y.; Jumper, J.; Kohli, P.; and Kelley, D.~R.
  2021{\natexlab{a}}.
\newblock Effective Gene Expression Prediction from Sequence by Integrating
  Long-Range Interactions.
\newblock \emph{Nature Methods}, 18: 1196--1203.

\bibitem[{Avsec et~al.(2021{\natexlab{b}})Avsec, Weilert, Shrikumar, Krueger,
  Alexandari, Dalal, Fropf, McAnany, Gagneur, Kundaje, and
  Zeitlinger}]{avsec2021bpnet}
Avsec, {\v{Z}}.; Weilert, M.; Shrikumar, A.; Krueger, S.; Alexandari, A.;
  Dalal, K.; Fropf, R.; McAnany, C.; Gagneur, J.; Kundaje, A.; and Zeitlinger,
  J. 2021{\natexlab{b}}.
\newblock Base-Resolution Models of Transcription-Factor Binding Reveal Soft
  Motif Syntax.
\newblock \emph{Nature Genetics}, 53: 354--366.

\bibitem[{Beer and Tavazoie(2004)}]{beer2004predicting}
Beer, M.~A.; and Tavazoie, S. 2004.
\newblock Predicting gene expression from sequence.
\newblock \emph{Cell}, 117(2): 185--198.

\bibitem[{Bommasani et~al.(2021)Bommasani, Hudson, Adeli, Altman, Arora, von
  Arx, Bernstein, Bohg, Bosselut, Brunskill et~al.}]{bommasani2021foundation}
Bommasani, R.; Hudson, D.~A.; Adeli, E.; Altman, R.; Arora, S.; von Arx, S.;
  Bernstein, M.~S.; Bohg, J.; Bosselut, A.; Brunskill, E.; et~al. 2021.
\newblock On the Opportunities and Risks of Foundation Models.
\newblock \emph{arXiv preprint arXiv:2108.07258}.

\bibitem[{Bostrom and Durrett(2020)}]{bostrom2020bpe}
Bostrom, K.; and Durrett, G. 2020.
\newblock Byte Pair Encoding Is Suboptimal for Language Model Pretraining.
\newblock In \emph{Findings of EMNLP}, 4617--4624.

\bibitem[{Brown et~al.(2020)Brown, Mann, Ryder, Subbiah, Kaplan, Dhariwal,
  Neelakantan, Shyam, Sastry, Askell et~al.}]{brown2020language}
Brown, T.~B.; Mann, B.; Ryder, N.; Subbiah, M.; Kaplan, J.; Dhariwal, P.;
  Neelakantan, A.; Shyam, P.; Sastry, G.; Askell, A.; et~al. 2020.
\newblock Language Models are Few-Shot Learners.
\newblock In \emph{Advances in Neural Information Processing Systems},
  volume~33, 1877--1901.

\bibitem[{Dalla-Torre et~al.(2025)Dalla-Torre, Gonzalez, Mendoza-Revilla,
  Lopez~Carranza, Grzywaczewski, Oteri, Dallago, Trop, de~Almeida, Sirelkhatim
  et~al.}]{dallatorre2025nucleotide}
Dalla-Torre, H.; Gonzalez, L.; Mendoza-Revilla, J.; Lopez~Carranza, N.;
  Grzywaczewski, A.~H.; Oteri, F.; Dallago, C.; Trop, E.; de~Almeida, B.~P.;
  Sirelkhatim, H.; et~al. 2025.
\newblock Nucleotide Transformer: Building and Evaluating Robust Foundation
  Models for Human Genomics.
\newblock \emph{Nature Methods}, 22: 287--297.

\bibitem[{Devlin et~al.(2019)Devlin, Chang, Lee, and
  Toutanova}]{devlin2019bert}
Devlin, J.; Chang, M.-W.; Lee, K.; and Toutanova, K. 2019.
\newblock {BERT}: Pre-training of Deep Bidirectional Transformers for Language
  Understanding.
\newblock In \emph{Proceedings of NAACL-HLT}, 4171--4186.

\bibitem[{Grant, Bailey, and Noble(2011)}]{grant2011fimo}
Grant, C.~E.; Bailey, T.~L.; and Noble, W.~S. 2011.
\newblock {FIMO}: Scanning for Occurrences of a Given Motif.
\newblock \emph{Bioinformatics}, 27(7): 1017--1018.

\bibitem[{Gre{\v{s}}ov{\'a} et~al.(2023)Gre{\v{s}}ov{\'a}, Martinek,
  {\v{C}}ech{\'a}k, {\v{S}}ime{\v{c}}ek, and
  Alexiou}]{gresova2023genomicbenchmarks}
Gre{\v{s}}ov{\'a}, K.; Martinek, V.; {\v{C}}ech{\'a}k, D.; {\v{S}}ime{\v{c}}ek,
  P.; and Alexiou, P. 2023.
\newblock Genomic Benchmarks: A Collection of Datasets for Genomic Sequence
  Classification.
\newblock \emph{BMC Genomic Data}, 24: 25.

\bibitem[{Huang et~al.(2026)Huang, Zhou, Cui, Tapia-Pacheco, Amariuta, Li, and
  Shang}]{huang2026evolen}
Huang, N.; Zhou, X.; Cui, J.; Tapia-Pacheco, M.; Amariuta, T.; Li, Y.; and
  Shang, J. 2026.
\newblock {EvoLen}: Evolution-Guided Tokenization for {DNA} Language Model.
\newblock \emph{arXiv preprint arXiv:2604.08698}.

\bibitem[{Huang et~al.(2025)Huang, Gao, Xu, Zhao, Song, Gui, Lv, Chen, Cui, Li,
  and Wei}]{huang2025peace}
Huang, Y.; Gao, T.; Xu, H.; Zhao, Q.; Song, Y.; Gui, Z.; Lv, T.; Chen, H.; Cui,
  L.; Li, S.; and Wei, F. 2025.
\newblock {PEACE}: Empowering Geologic Map Holistic Understanding with {MLLMs}.
\newblock In \emph{Proceedings of the IEEE/CVF Conference on Computer Vision
  and Pattern Recognition}, 3899--3908.

\bibitem[{Ji et~al.(2021)Ji, Zhou, Liu, and Davuluri}]{ji2021dnabert}
Ji, Y.; Zhou, Z.; Liu, H.; and Davuluri, R.~V. 2021.
\newblock {DNABERT}: Pre-trained Bidirectional Encoder Representations from
  Transformers Model for {DNA}-Language in Genome.
\newblock \emph{Bioinformatics}, 37(15): 2112--2120.

\bibitem[{Kelley et~al.(2018)Kelley, Reshef, Bileschi, Belanger, McLean, and
  Snoek}]{kelley2018basenji}
Kelley, D.~R.; Reshef, Y.~A.; Bileschi, M.; Belanger, D.; McLean, C.~Y.; and
  Snoek, J. 2018.
\newblock Sequential Regulatory Activity Prediction across Chromosomes with
  Convolutional Neural Networks.
\newblock \emph{Genome Research}, 28(5): 739--750.

\bibitem[{Kim et~al.(2026)Kim, Shin, Kim, Jung, Lee, Lee, Ahn, and
  Han}]{kim2026dnachunker}
Kim, T.; Shin, J.; Kim, H.; Jung, Y.; Lee, J.; Lee, W.-C.; Ahn, S.; and Han, I.
  2026.
\newblock {DNACHUNKER}: Learnable Tokenization for {DNA} Language Models.
\newblock \emph{arXiv preprint arXiv:2601.03019}.

\bibitem[{Kovachki et~al.(2023)Kovachki, Li, Liu, Azizzadenesheli,
  Bhattacharya, Stuart, and Anandkumar}]{kovachki2023neuraloperator}
Kovachki, N.; Li, Z.; Liu, B.; Azizzadenesheli, K.; Bhattacharya, K.; Stuart,
  A.; and Anandkumar, A. 2023.
\newblock Neural Operator: Learning Maps Between Function Spaces with
  Applications to {PDEs}.
\newblock \emph{Journal of Machine Learning Research}, 24(89): 1--97.

\bibitem[{Kudo and Richardson(2018)}]{kudo2018sentencepiece}
Kudo, T.; and Richardson, J. 2018.
\newblock {SentencePiece}: A Simple and Language Independent Subword Tokenizer
  and Detokenizer for Neural Text Processing.
\newblock In \emph{Proceedings of EMNLP: System Demonstrations}, 66--71.

\bibitem[{Li et~al.(2026)Li, Yu, Wang, Liu, Yu, Zhou, Yang, Guo, Zhang, and
  Li}]{li2026mergedna}
Li, S.; Yu, K.; Wang, A.; Liu, Z.; Yu, C.; Zhou, J.; Yang, Q.; Guo, Y.; Zhang,
  X.; and Li, S.~Z. 2026.
\newblock {MergeDNA}: Context-Aware Genome Modeling with Dynamic Tokenization
  Through Token Merging.
\newblock In \emph{Proceedings of the AAAI Conference on Artificial
  Intelligence}, volume~40, 668--676.

\bibitem[{Mudge et~al.(2025)Mudge, Carbonell-Sala, Diekhans, Gonzalez~Martinez,
  Hunt, Jungreis, Loveland et~al.}]{mudge2025gencode}
Mudge, J.~M.; Carbonell-Sala, S.; Diekhans, M.; Gonzalez~Martinez, J.; Hunt,
  T.; Jungreis, I.; Loveland, J.~E.; et~al. 2025.
\newblock {GENCODE} 2025: Reference Gene Annotation for Human and Mouse.
\newblock \emph{Nucleic Acids Research}, 53(D1): D966--D975.

\bibitem[{Nguyen et~al.(2023)Nguyen, Poli, Faizi, Thomas, Birch-Sykes, Wornow,
  Patel, Rabideau, Massaroli, Bengio, Ermon, Baccus, and
  R{\'e}}]{nguyen2023hyenadna}
Nguyen, E.; Poli, M.; Faizi, M.; Thomas, A.; Birch-Sykes, C.; Wornow, M.;
  Patel, A.; Rabideau, C.; Massaroli, S.; Bengio, Y.; Ermon, S.; Baccus, S.~A.;
  and R{\'e}, C. 2023.
\newblock {HyenaDNA}: Long-Range Genomic Sequence Modeling at Single Nucleotide
  Resolution.
\newblock \emph{arXiv preprint arXiv:2306.15794}.

\bibitem[{Ovek~Baydar et~al.(2026)Ovek~Baydar, Rauluseviciute, Aronsen,
  Blanc-Mathieu, Bonthuis, de~Beukelaer et~al.}]{baydar2026jaspar}
Ovek~Baydar, D.; Rauluseviciute, I.; Aronsen, D.~R.; Blanc-Mathieu, R.;
  Bonthuis, I.; de~Beukelaer, H.; et~al. 2026.
\newblock {JASPAR} 2026: Expansion of Transcription Factor Binding Profiles and
  Integration of Deep Learning Models.
\newblock \emph{Nucleic Acids Research}, 54(D1): D184--D193.

\bibitem[{Patel et~al.(2024)Patel, Singhal, Wang, Pampari, Kasowski, and
  Kundaje}]{patel2024darteval}
Patel, A.; Singhal, A.; Wang, A.; Pampari, A.; Kasowski, M.; and Kundaje, A.
  2024.
\newblock {DART-Eval}: A Comprehensive {DNA} Language Model Evaluation
  Benchmark on Regulatory {DNA}.
\newblock In \emph{Advances in Neural Information Processing Systems},
  volume~37.

\bibitem[{Qiao et~al.(2024)Qiao, Ye, Ren, Bai, Liang, Ma, Dong, and
  Ouyang}]{qiao2024mxdna}
Qiao, L.; Ye, P.; Ren, Y.; Bai, W.; Liang, C.; Ma, X.; Dong, N.; and Ouyang, W.
  2024.
\newblock Model Decides How to Tokenize: Adaptive {DNA} Sequence Tokenization
  with {MxDNA}.
\newblock In \emph{Advances in Neural Information Processing Systems},
  volume~37.

\bibitem[{Radford et~al.(2021)Radford, Kim, Hallacy, Ramesh, Goh, Agarwal,
  Sastry, Askell, Mishkin, Clark et~al.}]{radford2021clip}
Radford, A.; Kim, J.~W.; Hallacy, C.; Ramesh, A.; Goh, G.; Agarwal, S.; Sastry,
  G.; Askell, A.; Mishkin, P.; Clark, J.; et~al. 2021.
\newblock Learning Transferable Visual Models from Natural Language
  Supervision.
\newblock In \emph{Proceedings of the 38th International Conference on Machine
  Learning}, volume 139 of \emph{Proceedings of Machine Learning Research},
  8748--8763.

\bibitem[{Raissi, Perdikaris, and Karniadakis(2019)}]{raissi2019pinn}
Raissi, M.; Perdikaris, P.; and Karniadakis, G.~E. 2019.
\newblock Physics-Informed Neural Networks: A Deep Learning Framework for
  Solving Forward and Inverse Problems Involving Nonlinear Partial Differential
  Equations.
\newblock \emph{Journal of Computational Physics}, 378: 686--707.

\bibitem[{Rust et~al.(2021)Rust, Pfeiffer, Vuli{\'c}, Ruder, and
  Gurevych}]{rust2021tokenizer}
Rust, P.; Pfeiffer, J.; Vuli{\'c}, I.; Ruder, S.; and Gurevych, I. 2021.
\newblock How Good Is Your Tokenizer? On the Monolingual Performance of
  Multilingual Language Models.
\newblock In \emph{Proceedings of ACL-IJCNLP}, 3118--3135.

\bibitem[{Sanchez-Gonzalez et~al.(2020)Sanchez-Gonzalez, Godwin, Pfaff, Ying,
  Leskovec, and Battaglia}]{sanchezgonzalez2020gns}
Sanchez-Gonzalez, A.; Godwin, J.; Pfaff, T.; Ying, R.; Leskovec, J.; and
  Battaglia, P. 2020.
\newblock Learning to Simulate Complex Physics with Graph Networks.
\newblock In \emph{Proceedings of the 37th International Conference on Machine
  Learning}, volume 119 of \emph{Proceedings of Machine Learning Research},
  8459--8468.

\bibitem[{Schiff et~al.(2024)Schiff, Kao, Gokaslan, Dao, Gu, and
  Kuleshov}]{schiff2024caduceus}
Schiff, Y.; Kao, C.-H.; Gokaslan, A.; Dao, T.; Gu, A.; and Kuleshov, V. 2024.
\newblock Caduceus: Bi-Directional Equivariant Long-Range {DNA} Sequence
  Modeling.
\newblock In \emph{Proceedings of the 41st International Conference on Machine
  Learning}, volume 235 of \emph{Proceedings of Machine Learning Research},
  43632--43648.

\bibitem[{Sennrich, Haddow, and Birch(2016)}]{sennrich2016bpe}
Sennrich, R.; Haddow, B.; and Birch, A. 2016.
\newblock Neural Machine Translation of Rare Words with Subword Units.
\newblock In \emph{Proceedings of ACL}, 1715--1725.

\bibitem[{Shrikumar et~al.(2018)Shrikumar, Tian, Avsec, Shcherbina, Banerjee,
  Sharmin, Nair, and Kundaje}]{shrikumar2018technical}
Shrikumar, A.; Tian, K.; Avsec, {\v{Z}}.; Shcherbina, A.; Banerjee, A.;
  Sharmin, M.; Nair, S.; and Kundaje, A. 2018.
\newblock Technical Note on Transcription Factor Motif Discovery from
  Importance Scores (TF-MoDISco) version 0.5.6.5.
\newblock \emph{arXiv preprint arXiv:1811.00416}.

\bibitem[{Stormo(2013)}]{stormo2013specificity}
Stormo, G.~D. 2013.
\newblock Modeling the Specificity of Protein--{DNA} Interactions.
\newblock \emph{Quantitative Biology}, 1(2): 115--130.

\bibitem[{Sundararajan, Taly, and Yan(2017)}]{sundararajan2017axiomatic}
Sundararajan, M.; Taly, A.; and Yan, Q. 2017.
\newblock Axiomatic Attribution for Deep Networks.
\newblock In \emph{Proceedings of the 34th International Conference on Machine
  Learning}, volume~70 of \emph{Proceedings of Machine Learning Research},
  3319--3328.

\bibitem[{Vorontsov et~al.(2024)}]{vorontsov2024hocomoco}
Vorontsov, I.~E.; et~al. 2024.
\newblock {HOCOMOCO} in 2024: A Rebuild of the Curated Collection of Binding
  Models for Human and Mouse Transcription Factors.
\newblock \emph{Nucleic Acids Research}, 52(D1): D154--D163.

\bibitem[{Wang et~al.(2023)Wang, Fu, Du, Gao, Huang, Liu, Chandak, Liu,
  Van~Katwyk, Deac et~al.}]{wang2023scientificdiscovery}
Wang, H.; Fu, T.; Du, Y.; Gao, W.; Huang, K.; Liu, Z.; Chandak, P.; Liu, S.;
  Van~Katwyk, P.; Deac, A.; et~al. 2023.
\newblock Scientific Discovery in the Age of Artificial Intelligence.
\newblock \emph{Nature}, 620: 47--60.

\bibitem[{Xue et~al.(2022)Xue, Barua, Constant, Al-Rfou, Narang, Kale, Roberts,
  and Raffel}]{xue2022byt5}
Xue, L.; Barua, A.; Constant, N.; Al-Rfou, R.; Narang, S.; Kale, M.; Roberts,
  A.; and Raffel, C. 2022.
\newblock {ByT5}: Towards a Token-Free Future with Pre-trained Byte-to-Byte
  Models.
\newblock \emph{Transactions of the Association for Computational Linguistics},
  10: 291--306.

\bibitem[{Zhou and Troyanskaya(2015)}]{zhou2015deepsea}
Zhou, J.; and Troyanskaya, O.~G. 2015.
\newblock Predicting Effects of Noncoding Variants with Deep Learning-Based
  Sequence Model.
\newblock \emph{Nature Methods}, 12: 931--934.

\bibitem[{Zhou et~al.(2025)Zhou, Wang, Shang, and
  Li}]{zhou2025dnamotiftokenizer}
Zhou, X.; Wang, Z.; Shang, J.; and Li, Y.~E. 2025.
\newblock {DNAMotifTokenizer}: Towards Biologically Informed Tokenization of
  Genomic Sequences.
\newblock \emph{arXiv preprint arXiv:2512.17126}.

\bibitem[{Zhou et~al.(2024)Zhou, Ji, Li, Dutta, Davuluri, and
  Liu}]{zhou2024dnabert2}
Zhou, Z.; Ji, Y.; Li, W.; Dutta, P.; Davuluri, R.~V.; and Liu, H. 2024.
\newblock {DNABERT}-2: Efficient Foundation Model and Benchmark for
  Multi-Species Genomes.
\newblock In \emph{International Conference on Learning Representations}.

\end{thebibliography}

\end{document}